\documentclass[12pt]{article}
\usepackage{cite}
\usepackage{enumerate}
\usepackage{ifpdf}
\usepackage{ae}
\usepackage[T1]{fontenc}
\usepackage[ansinew]{inputenc}
\usepackage{amsmath}
\usepackage{relsize}
\usepackage{amssymb}
\usepackage{tikz}
\usepackage{graphicx}
\usepackage{color}
\definecolor{darkblue}{cmyk}{0.9,0.9,0,0}
\definecolor{darkgreen}{rgb}{0,0.55,0}
\usepackage[colorlinks=true,linkcolor=darkblue,citecolor=darkblue,urlcolor=darkblue]{hyperref}
\usepackage{epsfig}
\usepackage{epstopdf}
\usepackage{graphicx}
\DeclareMathOperator*{\dDisc}{dDisc}

\makeatletter
\long\def\@makecaption#1#2{
  \vskip\abovecaptionskip
  \sbox\@tempboxa{{\captionfonts #1: #2}}
  \ifdim \wd\@tempboxa >\hsize
    {\captionfonts #1: #2\par}
  \else
    \hbox to\hsize{\hfil\box\@tempboxa\hfil}
  \fi
  \vskip\belowcaptionskip}
\makeatother

\newcommand{\beq}{\begin{equation}}
\newcommand{\eeq}{\end{equation}}
\newcommand{\beqy} {\begin{eqnarray}}
\newcommand{\eeqy} {\end{eqnarray}}
\newcommand{\bsmat}{\begin{smallmatrix}}
\newcommand{\esmat}{\end{smallmatrix}}
\newcommand{\bmat}{\begin{matrix}}
\newcommand{\emat}{\end{matrix}}

\def\({\left(}
\def\){\right)}
\def\[{\left[}
\def\]{\right]}

\def\<{\langle}
\def\>{\rangle}

\usepackage{varioref}
\usepackage{makeidx}
\makeindex

\usepackage[english]{babel}
\usepackage{caption}
\usepackage{subfig}

\usepackage{tabularx}
\begin{document}

\thispagestyle{empty}

\renewcommand{\thefootnote}{\fnsymbol{footnote}}
\setcounter{page}{1}
\setcounter{footnote}{0}
\setcounter{figure}{0}
\begin{titlepage}

\begin{center}

\vskip 2.3 cm 

\vskip 5mm

{\Large \bf
The Analytic Bootstrap for Large $N$}

{\Large \bf Chern-Simons Vector Models
}
\vskip 0.5cm

\vskip 15mm

\centerline{Ofer Aharony$^{a}$, Luis F. Alday$^{b}$, Agnese Bissi$^{c}$ and  Ran Yacoby$^a$}
\bigskip
\centerline{\it $^{a}$ Department of Particle Physics and Astrophysics, Weizmann Institute of Science,}
\centerline{\it Rehovot 7610001, Israel}
\smallskip
\centerline{\it $^{b}$  Mathematical Institute, University of Oxford,} 
\centerline{\it  Andrew Wiles Building, Radcliffe Observatory Quarter,}
\centerline{\it Woodstock Road, Oxford, OX2 6GG, UK}
\smallskip
\centerline{\it $^{c}$ Department of Physics and Astronomy, Uppsala University,}
\centerline{\it Box 516, SE-75120 Uppsala, Sweden}

\end{center}

\vskip 2 cm

\begin{abstract}
\noindent 
Three-dimensional Chern-Simons vector models display an approximate higher spin symmetry in the large $N$ limit. Their single-trace operators consist of a tower of weakly broken currents, as well as a scalar $\sigma$ of approximate twist $1$ or $2$. We study the consequences of crossing symmetry for the four-point correlator of $\sigma$ in a $1/N$ expansion, using analytic bootstrap techniques. To order $1/N$ we show that crossing symmetry fixes the contribution from the tower of currents, providing an alternative derivation of well-known results by Maldacena and Zhiboedov. When $\sigma$ has twist $1$ its OPE receives a contribution from the exchange of $\sigma$ itself with an arbitrary coefficient, due to the existence of a marginal sextic coupling. We develop the machinery to determine the corrections to the OPE data of double-trace operators due to this, and to similar exchanges. This in turns allows us to fix completely the correlator up to three known truncated solutions to crossing. We then proceed to study the problem to order $1/N^2$. We find that crossing implies the appearance of odd-twist double-trace operators, and calculate their OPE coefficients in a large spin expansion. Also, surprisingly, crossing at order $1/N^2$, implies non-trivial $O(1/N)$ anomalous dimensions for even-twist double-trace operators, even though such contributions do not appear in the four-point function at order $1/N$ (in the case where there is no scalar exchange). We argue that this phenomenon arises due to operator mixing. Finally, we analyse the bosonic vector model with a sextic coupling without gauge interactions, and determine the order $1/N^2$ corrections to the dimensions of twist-$2$ double-trace operators. 

\end{abstract}

\end{titlepage}


\setcounter{page}{1}
\renewcommand{\thefootnote}{\arabic{footnote}}
\setcounter{footnote}{0}

 \def\nref#1{{(\ref{#1})}}

\section{Introduction and summary}

In the last few years, since the pioneering work of \cite{Rattazzi:2008pe}, great progress has been made in studying conformal field theories in $d>2$ dimensions using the conformal bootstrap \cite{Polyakov:1974gs,Ferrara:1973yt,Mack:1975jr}, both numerically and analytically. Analytic methods are most powerful when applied to conformal field theories in specific limits, parametrised by a small parameter. Two limits that have been studied are the large $N$ limit and weak coupling limits. The large $N$ limit maps to a perturbative expansion of a holographically dual bulk description, while weak coupling limits correspond to the ordinary perturbative expansion in field theory, in which it has approximate higher-spin symmetry with an infinite tower of almost-conserved currents.

In $d=3$ Chern-Simons-matter theories in the 't Hooft limit (and only in these theories) both expansions coexist -- these theories have a large $N$ expansion but also an approximate higher-spin symmetry in the large $N$ limit, even when they are not weakly coupled \cite{Aharony:2011jz,Giombi:2011kc}. It is the aim of this paper to develop an analytic bootstrap approach to analyse these theories. At leading order in the $1/N$ expansion, many computations in these theories can be performed (for any value of the couplings), following \cite{Giombi:2011kc}, by quantum field theory methods. As we will see below, bootstrap methods can alternatively be used to determine the  simplest four-point functions to this order, up to a finite, small, number of parameters (this was also shown in \cite{Turiaci:2018nua}). At the next order in $1/N$ very little is known from the field theory side, but we will show that bootstrap methods do enable some computations to be performed. In generic Chern-Simons-matter theories we will see that their power is limited because of operator mixings, but in the specific case of $\phi^6$ theories with no Chern-Simons coupling, more information can be obtained. The methods that we develop should be useful also for various other theories.

Three dimensional QFTs with enhanced higher-spin symmetry in the large $N$ limit were classified in \cite{Maldacena:2012sf}. They are:\footnote{In the language of \cite{Maldacena:2012sf} we are looking here only at theories which have a single conserved energy-momentum tensor in the large $N$ limit, which corresponds to having a single flavor; the generalization to more than one matter field is interesting but is beyond the scope of this paper.}
\begin{itemize}
\item
Scalar: A theory of $N$ complex scalar fields $\phi_i$ ($i=1,\cdots,N$) with a $U(N)$ global symmetry and an interaction
\beq \label{lambdasix}
{\cal L}_{\mathrm{int.}}=-\frac{\lambda_6}{6N^2} (|\vec{\phi}|^2)^3 \,,
\eeq
at the tri-critical fixed point where the mass and the $(|\vec{\phi}|^2)^2$ coupling are tuned to zero;
\item
CS-Scalar: The $U(N)_k$ Chern-Simons (CS) theory coupled to a single complex scalar field in the fundamental representation, with an 't~Hooft coupling $\lambda \equiv N/k$, with a $\lambda_6$ coupling as in \eqref{lambdasix}, and at a similar tri-critical fixed point;
\item
CS-Fermion: The $U(N)_k$ Chern-Simons theory coupled to a single Dirac fermion in the fundamental representation, with an 't~Hooft coupling $\lambda \equiv N/k$, at the fixed point where the fermion mass is tuned to zero.
\end{itemize}

The first theory, when restricted to the singlet sector of $U(N)$, may be viewed as a special case of the second one, with $\lambda=0$; note that for $\lambda > 0$ there is nothing special about $\lambda_6=0$, since $\lambda_6$ is generated at order $1/N$ in any case. All of these theories have also versions with real scalar fields or fermions and $SO(N)$ instead of $U(N)$; there are also versions where $SU(N)$ instead of $U(N)$ is gauged, whose correlation functions are the same to the order in $1/N$ that we work in.

The first two theories are conformal at leading order in the large $N$ expansion, but have a beta function for $\lambda_6$ at order $1/N$, and we expect them to be conformal only for specific values of this coupling (for large enough values of $N$ there is at least one zero of the beta function for any $\lambda$ \cite{Aharony:2018pjn}). 

The CS-Scalar theory is believed to be the same (at least for large enough values of $N$) as CS coupled to critical fermions, and the CS-Fermion theory the same as CS coupled to critical scalars \cite{Giombi:2011kc,Aharony:2012nh,Aharony:2015mjs}, so there is no need to discuss the latter two theories separately.

In all these theories there is a tower of single-trace operators $J_s$, with $s=1,2,\cdots$, of twist $\Delta-s=1+O(1/N)$, which become conserved currents in the large $N$ limit. In the first two theories there is a scalar $J_0$ of twist $1+O(1/N)$, and in the third theory a scalar $\sigma$ (which we will also denote by $J_0$) of twist $2+O(1/N)$. The single-trace operators are bilinears in the scalars/fermions with some insertions of (covariant) derivatives acting on them; in the first two theories we have schematically $J_s = \vec{\phi}^{\dagger} D_{\mu_1} \cdots D_{\mu_s} \vec{\phi}$. The other operators with finite dimensions in the large $N$ limit are multi-trace operators, which at infinite $N$ are products of the single-trace operators. Double-trace operators of spin $s$ will be denoted by $[J_{s_1} J_{s_2}]_{n,s}$. The operators $[J_0 J_0]_{n,s}$ are uniquely labeled by their twist $2+2n$ and their spin $s$.

Our strategy will be to consider the simplest $4$-point correlator of the theory 
\begin{equation}
\langle J_0(x_1) J_0(x_2) J_0(x_3) J_0(x_4) \rangle \,,
\end{equation}
and to analyse the constraints of crossing symmetry in a large $N$ expansion. For 
simplicity we will normalize the single-trace operators so that their two-point functions are of order one at large $N$. The general form of the $J_0 \times J_0$ OPE in a large $N$ expansion was discussed in \cite{Heemskerk:2009pn}. Schematically  we can write
\begin{equation} 
J_0 \times J_0 = 1 + [J_0 J_0]_{n,s} + \frac{1}{\sqrt{N}} J_s + O(\frac{1}{N}) \,.
\end{equation}
 At $O(1)$ the contribution to the four-point function arises only from the disconnected diagrams (products of two-point functions). Expanding this in conformal blocks, the identity operator contributes in one of the diagrams, and otherwise only the $[J_0 J_0]_{n,s}$ operators contribute to the OPE, with known OPE coefficients squared $a^{(0)}_{n,s}$. At order $1/N$ there are contributions from single-trace intermediate states $J_s$, depending on their OPE coefficients squared $a_{s}^{\mathrm{st},(0)}$ which are known for any $\lambda$ and $\lambda_6$ \cite{Maldacena:2012sf,Aharony:2012nh,GurAri:2012is}, and also from the same double-trace intermediate states, proportional either to their anomalous dimensions at order $1/N$, $\gamma_{n,s}^{(1)}$, or to the order $1/N$ correction to their OPE coefficients squared, $a_{n,s}^{(1)}$. 

At order $1/N$ it is not known how to directly compute the four-point function of the CS theories by field theory methods, though it is known in some specific kinematic regimes \cite{Bedhotiya:2015uga,Turiaci:2018nua,Yacoby:2018yvy}. However, one can determine the correlation function almost uniquely just from bootstrap considerations. In general large $N$ theories, at this order the double-trace contributions are determined by the single-trace contributions, up to homogeneous solutions to the bootstrap equation, see \cite{Heemskerk:2009pn,Alday:2016njk}; moreover, in large $N$ theories, only homogeneous solutions which vanish for $s>2$ are allowed \cite{Caron-Huot:2017vep, Alday:2017vkk,Simmons-Duffin:2017nub}, leaving $3$ possible solutions. As we will see, the contributions $a_{s}^{\mathrm{st},(0)}$ from single-trace operators can actually be fixed just by using crossing symmetry, at least for $s\geq 2$, providing an alternative derivation of this known result \cite{Maldacena:2012sf}. In the CS-Fermion theory the single-trace $\langle J_0 J_0 J_s \rangle$ three-point functions for any $\lambda$ are proportional to those of the $\lambda=0$ theory, with the proportionality constant independent of $s$. Thus, the four-point function is also proportional to that of the free theory, up to the truncated solutions which may be shown to vanish by an explicit computation \cite{Turiaci:2018nua}. In the CS-Scalar theory the same is true for a specific value of $\lambda_6$ (depending on $\lambda$), so for that value the four-point function is again proportional to that of the free theory, up to truncated solutions which may be shown to vanish \cite{Yacoby:2018yvy}. The contributions at this order from the $\lambda_6$ coupling are easy to compute, since this coupling only changes the coefficient of the $J_0$ term in the single-trace sector of the OPE by a constant times $\lambda_6$. The full crossing-symmetric contribution to the four-point function due to $\lambda_6$ is then given by $\phi^3$-type exchange Witten diagrams in $AdS_4$, which can be explicitly written in terms of $\overline{D}$-functions.

In this paper we discuss what can be said about the four-point function and the related CFT data at order $1/N^2$, using the conformal bootstrap. More precisely, we will use the analytic approach of \cite{Alday:2016njk}, initiated in \cite{Fitzpatrick:2012yx, Komargodski:2012ek} in the non-perturbative regime, and in \cite{Alday:2013cwa} for theories with weakly broken higher-spin symmetry. This approach has already been applied to large $N$ theories, see \cite{Aharony:2016dwx, Alday:2017gde, Alday:2017xua}, but the presence of an infinite tower of new intermediate operators to order $1/N$ and the fact that we are in $d=3$ make the analysis quite different from previous cases, and we develop new analytic bootstrap methods that are required for this. From the point of view of a dual bulk theory, $1/N^2$ corresponds to one-loop order; these theories were conjectured in \cite{Klebanov:2002ja,Sezgin:2003pt,Giombi:2009wh,Giombi:2011kc,Chang:2012kt} to be holographically dual to Vasiliev-type higher spin gravity theories on $AdS_4$ \cite{Vasiliev:1992av}, so our results constrain one-loop computations in these theories (about which nothing is presently known).

\subsection{Summary of results}

In theories with only one single-trace operator, once the theory is known at order $1/N$, one can also compute the correlation functions at order $1/N^2$, up to the same freedom in adding truncated solutions, see  \cite{Aharony:2016dwx}; the main input that is needed for this is the value of $(\gamma_{n,s}^{(1)})^2$. In the dual bulk theory, this is essentially the statement that once a tree-level field theory is given, the one-loop amplitudes are determined, up to possible extra local interactions arising at one-loop order. However, in our case the situation is more complicated because of operator mixings. Generic four-point functions $\langle J_{s_1} J_{s_2} J_{s_3} J_{s_4} \rangle$ are non-zero at order $1/N$, and this leads to a mixing of the double-trace operators at this order. The actual eigenvectors of the anomalous dimension matrix are linear combinations $\Sigma_i$ of the $[J_s J_{s'}]$ operators, with coefficients that are generally of order $1$, and thus the anomalous dimension $\gamma_{n,s}^{(1)}$ mentioned above (that appears in the four-point function at order $1/N$) is actually a linear combination of the anomalous dimensions of all the $\Sigma_i$ operators that contain $[J_0 J_0]_{n,s}$ (we will call this the ``average value of $\gamma_{n,s}^{(1)}$''). But if this is all we know then we cannot compute the same linear combination of the anomalous dimensions squared, needed to go to the next order; and in order to determine the mixing we need to know all $\langle J_{s_1} J_{s_2} J_{s_3} J_{s_4} \rangle$ correlation functions up to order $1/N$, and not just that of $J_0$. Unfortunately almost nothing is known about these general correlation functions, whose bootstrap analysis involves spinning conformal blocks.\footnote{A very similar mixing problems arises in ${\cal N}=4$ SYM to order $1/N^4$, but in that case the mixing can be solved by resorting only to additional scalar correlators, see \cite{Alday:2017xua,Aprile:2017bgs}. } In the bulk language, we need to know all the tree-level vertices that can appear in the one-loop diagrams in order to compute a correlation function at one-loop order, and not just a specific one. Note that in the Scalar theory there is no such mixing for the $[J_0 J_0]_{n,s}$ operators for large $\lambda_6$, since the other four-point functions involving two $J_0$ operators have a different dependence on $\lambda_6$, so this issue does not arise at order $\lambda_6^2$ (but only at lower orders).

Because of this issue, we will be able to make less progress in understanding the $1/N^2$ corrections than in theories with no mixing, but we will still be able to make quantitative claims about averages. At this order, it is still true that only single-trace and double-trace operators appear in the OPE, but in general all double-trace operators will appear (and not just the ones that have an overlap with $[J_0 J_0]$). There are various contributions to the conformal block expansion of $\langle J_0 J_0 J_0 J_0 \rangle$ at this order:

\begin{itemize}

\item There are contributions proportional to the order $1/N$ single-trace anomalous dimensions $\gamma_{s}^{\mathrm{st},(1)}$, and to the order $1/N$ corrections to their OPE coefficients squared $a_s^{\mathrm{st},(1)}$. The anomalous dimensions at order $1/N$ for $s>0$ were computed in \cite{Giombi:2016zwa}  for any $\lambda$, and they are independent of $\lambda_6$ (in particular they vanish in the Scalar theory). However, $\gamma_0^{\mathrm{st},(1)}$ is not known (except at $\lambda=0$ where it vanishes for any $\lambda_6$). In the CS-Fermion theory its value is not relevant for the intermediate operators at this order, since $a_0^{\mathrm{st},(0)}$ vanishes, but this is not true in the CS-Scalar theory. In both theories,  $\gamma_0^{\mathrm{st},(1)}$ appears in the four-point function through the dimension of the external operators, leading to a contribution proportional to $\gamma_0^{\mathrm{st},(1)}$ times the four-point function at order $1/N$. In the Scalar theory we will argue that the $a_s^{\mathrm{st},(1)}$ with $s>0$ all vanish, using the fact that in this theory the $J_s$ are the only odd-twist operators that appear up to order $1/N^2$ (at higher orders, triple-trace operators also appear). In the other theories the $a_s^{\mathrm{st},(1)}$  are not known.

\item There are contributions involving the double-trace operators that overlap (at leading order) with $[J_0 J_0]$. As for $\gamma_{n,s}^{(1)}$ and $a_{n,s}^{(1)}$, the four-point function at this order involves averages over these operators with appropriate weights; the computation at order $1/N^2$ involves the averages of  $(\gamma_{n,s}^{(1)})^2$, $a_{n,s}^{(1)} \gamma_{n,s}^{(1)}$, $a_{n,s}^{(2)}$ and $\gamma_{n,s}^{(2)}$. In the Scalar theory where there is no mixing at order $\lambda_6^2$ we can compute the $\gamma_{n,s}^{(2)}$ at leading order (for large $\lambda_6$) using the methods of \cite{Alday:2016njk,Caron-Huot:2017vep}, since in this theory the known $(\gamma_{n,s}^{(1)})^2$'s are the only sources of double discontinuities. We perform this computation for the leading twist operators and obtain (for large $\lambda_6$)
\begin{equation}
\gamma_{0,s} =- \frac{4 a_0}{N} \frac{1}{2s+1}  - \frac{2 a_0^2}{N^2}\frac{\pi  \Gamma \left(\frac{s+1}{2}\right)^2}{(2s+1)\Gamma \left(\frac{s+2}{2}\right)^2} + O(\frac{\lambda_6^3}{N^2}, \frac{1}{N^3})\,,~~~~~~\text{Scalar theory}
\end{equation}
where $a_0=\frac{\lambda_6(\lambda_6-256)}{2^{12}}+O(1/N)$. In the other theories it is not clear how to compute the $\gamma_{n,s}^{(2)}$  without resolving the mixing. Note that even though in some of the other theories the averages $(\gamma_{n,s}^{(1)})$ vanish, the average of their squares, which is related by crossing to the $\gamma_s^{\mathrm{st},(1)}$'s, cannot vanish, showing that mixing must occur; we will be able to obtain some constraints on these averages-of-squares.

\item Finally, at this order we start getting a contribution also from double-trace operators which have no overlap with $[J_0 J_0]$; since these operators only start contributing at this order, their anomalous dimensions do not show up, but only their OPE coefficients squared $a_{other}^{(0)}$ (which are of order $1/N^2$). Naively all double-trace operators in these theories, except for $[J_0 J_s]$ in the CS-Fermion theory, have even twist. However, using an $\epsilon$-symbol, one can construct double-trace operators of odd twist from other products of various single-trace operators. All of these first appear in the $J_0$ four-point function at this order, but they do not appear in the Scalar theory where parity does not allow it. In the CS-Fermion theory we will be able to obtain some information about the average $a_{\mathrm{other}}^{(0)}$'s of the odd-twist operators, since they are related by crossing to specific combinations of the known $\gamma_{s}^{\mathrm{st},(1)}$'s.

\end{itemize}

\subsection{Future directions}

In this paper we make some progress on computing the four-point functions of CS-matter theories at order $1/N^2$ and their various components. It would be nice to make further progress on this. In particular, computing additional four-point functions beyond those of the scalar $J_0$, and including their information, should determine the mixings between the various double-trace operators, and enable much more precise results to be obtained. It would be nice to understand what can be said about higher orders in the $1/N$ expansion, and about non-perturbative corrections to this expansion (say of order $e^{-N}$), which should also be consistent with crossing. 

Just from our procedure we cannot see the beta function of $\lambda_6$ in the Scalar and CS-Scalar theories; for values of $\lambda_6$ for which the beta function at order $1/N$ is non-zero, the three-point function $\langle J_0 J_0 J_0 \rangle$ will acquire a logarithmic scale-dependence which will affect the four-point function at order $1/N^2$, but in the large $N$ bootstrap approach this three-point function is an input to the computation that is not constrained by itself. Our results assume that there are no such logarithms, so they are valid at values of $\lambda_6$ for which the beta function vanishes. A finite $N$ bootstrap computation should be able to compute the 3-point function (and thus the allowed values of $\lambda_6$).

For the Scalar theory the fundamental field $\phi^i$ is also part of the spectrum. It would be interesting to study analytically the system of mixed correlators involving $J_0$ and $\phi^i$, and to use it to obtain further information.

It would be interesting to study various generalizations of our analysis, to theories with supersymmetry (in particular ${\cal N}=1$ or ${\cal N}=2$ supersymmetry), general theories with more fields (e.g. both scalars and fermions), theories with product gauge groups, etc.

\section{Analytic bootstrap for CS-matter theories}

\subsection{General strategy}
Our strategy will be to consider the four-point correlator of the simplest scalar operator of the CS-matter theories under consideration, which we will denote here as $J_0$. The dimension of this operator in the theories of interest to us is either $\Delta_{0} = 1+ O(1/N)$ or $\Delta_{0} = 2+ O(1/N)$, depending on the model under consideration, but in many places we will keep it as an arbitrary parameter in our equations. Conformal invariance implies the structure
\begin{equation}
\label{J0corr}
\langle J_0(x_1)J_0(x_2)J_0(x_3)J_0(x_4) \rangle = \frac{{\cal G}(u,v) }{x_{12}^{2\Delta_{0}}x_{34}^{2\Delta_{0}}}\,,
\end{equation}
where we have introduced conformal cross-ratios $u=\frac{x_{12}^2 x_{34}^2}{x_{13}^2 x_{24}^2},v=\frac{x_{14}^2 x_{23}^2}{x_{13}^2 x_{24}^2}$. An important property, to be used heavily in the present work, is that the correlator satisfies crossing symmetry
\begin{equation}\label{crossing}
v^{\Delta_{0}}{\cal G}(u,v) = u^{\Delta_{0}}{\cal G}(v,u) .
\end{equation}
Furthermore, it also admits a decomposition in conformal blocks
\begin{equation}
{\cal G}(u,v) =\sum_{\tau,s} a_{\tau,s} f_{\tau,s}(u,v)\,,
\end{equation}
where $f_{\tau,s}(u,v)$ are three-dimensional conformal blocks, related to the contribution in an OPE expansion of intermediate states with spin $s$ and twist $\tau=\Delta-s$. We find it convenient to use conventions in which 
\begin{equation}
f_{\tau,s}(u,v) = r_{\frac{\tau}{2}+s}r_{\frac{\tau}{2}} \tilde f_{\tau,s}(u,v),~~~r_h=\frac{\Gamma(h)^2}{\Gamma(2h)}\,,
\end{equation}
and $\tilde f_{\tau,s}(u,v)$ has the usual normalisation, namely $\tilde f_{\tau,s}(u,v)  \to u^{\tau/2}$ for small $u$ and $v \to 1$. Although the conformal blocks in three dimensions are not known in a closed form, they can be written as infinite sums of $SL(2,R)$ conformal blocks. This, together with their known properties, will be enough for our purposes. 

We will consider these correlators in a large $N$ expansion. To leading order the result reduces to that of generalized free fields
\begin{equation}\label{gff}
{\cal G}^{(0)}(u,v) = 1+ \left(\frac{u}{v}\right)^{\Delta_{0}} + u^{\Delta_{0}} \,,
\end{equation}
in which the intermediate operators are the identity operator and double-trace operators of spin $s$ and twist $2\Delta_{0}+2n$ (denoted as $[J_0 J_0]_{n,s}$). We will study $1/N$ corrections to this:
\begin{equation}
{\cal G}(u,v) = 1+ \left(\frac{u}{v}\right)^{\Delta_{0}} + u^{\Delta_{0}} +\frac{1}{N}  {\cal G}^{(1)}(u,v)+ \frac{1}{N^2}  {\cal G}^{(2)}(u,v)+\cdots\,.
\end{equation}
These corrections may arise from two sources. First, the twist and OPE coefficients of double-trace operators get corrected
\begin{eqnarray}\label{largenexp}
\tau_{n,s} &=& 2\Delta_{0} + 2n + \frac{1}{N} \gamma^{(1)}_{n,s} + \cdots\,,\\
a_{n,s} &=& a^{(0)}_{n,s} + \frac{1}{N} a^{(1)}_{n,s} + \cdots\,.
\end{eqnarray}
In addition, we will have new intermediate operators entering at order $1/N$. We will use symmetries to constrain both types of contributions. 

From \cite{Alday:2016njk,Caron-Huot:2017vep} it follows that the full CFT data can be reconstructed from the double-discontinuities, or enhanced singularities, of the correlator, up to a finite ambiguity which is easily characterised. We will use this fact together with crossing symmetry to constrain the spectrum of anomalous dimensions and OPE coefficients of these theories. 

There are two equivalent methods to reconstruct the CFT data from the enhanced singularities as $v \to 0$. The first is large spin perturbation theory (LSPT) \cite{Alday:2016njk}. The idea is to consider specific sums (denoted as twist conformal blocks (TCBs)) with prescribed enhanced-singularities. These sums are given by
\begin{equation}
H^{(m)}_\tau(u,v) \equiv \sum_{s=0,2,\cdots} \frac{f_{\tau,s}(u,v)}{J^{2m}}\,,
\end{equation}
where the conformal spin is defined as $J^2=\frac{1}{4} (2 s+\tau -2) (2 s+\tau )$. In appendix \ref{appTCB} it is shown how to compute the enhanced-singular part of $H^{(m)}_\tau(u,v)$ as a series expansion around small $u,v$ and for all $\tau$ and $m$. Given these functions, we can always express a given enhanced-singularity as linear combinations of $H^{(m)}_\tau(u,v)$ and their $\partial_\tau$ derivatives. This allows the reconstruction of the CFT data  as an expansion in inverse powers of the spin, to all orders. By enhanced-singularity we mean a behaviour that becomes singular after applying the Casimir operator a finite number of times; this includes terms like $1/v,\log^2 v$ and $v^{1/2}$. Equivalently, one can use the inversion formula in \cite{Caron-Huot:2017vep}, which gives the CFT data in terms of the double-discontinuity of the correlator. This can be seen as repackaging the whole series above, in terms of integrals involving hypergeometric functions. Note that the concepts of enhanced singularity and double-discontinuity are equivalent. 

We will focus on two distinct types of problems. In one problem, crossing symmetry maps an enhanced-singularity in both channels, for small $u,v$, to a contribution of the same type. In the second kind of problems, given a specific enhanced-singularity in one channel, we need to recover the CFT data corresponding to it. Let us start with a problem of the first type.  

\subsection{OPE coefficients of single-trace operators}\label{stope}
As a simple application of LSPT, let us consider the following problem. To leading order in $1/N$ the intermediate operators of the correlator \eqref{J0corr} are double-trace operators of even twist. To order $1/N$ single-trace operators $J_s$ also arise with a specific OPE coefficient. We would like to fix this OPE coefficient from crossing symmetry. In our derivation it will be important that no other operators of odd-twist appear at this order. The currents $J_s$ appear at order $1/N$ with their classical twist, so that their contribution to the correlator is given by\footnote{We will discuss separately the contribution from the spin-$0$ single-trace operator, which may also have twist one.}
\begin{equation}
 \left. {\cal G}^{(1)}(u,v) \right|_{\mathrm{st}}=\sum_{s=2,4,\cdots} a^{\mathrm{st},(0)}_{s}  f_{\tau=1,s}(u,v) \,.
\end{equation}
Next, we assume that the OPE coefficients squared $a^{\mathrm{st},(0)}_{s}$ admit an expansion 
\begin{equation}\label{largejexp}
a^{\mathrm{st},(0)}_{s} = \frac{1}{N} \sum_m \frac{\alpha_m}{J^{2m}} \,.
\end{equation}
The contribution from single-trace operators can then be written as a linear combination of the twist conformal blocks introduced above, with $\tau=1$:
\begin{equation}
 \left. {\cal G}^{(1)}(u,v) \right|_{\mathrm{st}}= \sum_m \alpha_m H^{(m)}_1(u,v)\,.
\end{equation}
Now comes an important point: the twist of the conserved currents is one. Furthermore, the structure of $H^{(m)}_1(u,v)$ is such that, for integer $m$, it has enhanced singularities that are given by an expansion around small $u,v$ involving half-integer powers of $u$ and $v$.\footnote{Recall that enhanced singularities can come from negative powers of $v$, non-integer positive powers, or terms of the form $\log^n(v)$ with $n \geq 2$. A more detailed analysis below shows that actually $m$ in \eqref{largejexp} has to be integer for integer $\Delta_{0}$. 
	
} If no other operators of odd twist appear to this order, and for integer classical $\Delta_{0}$,  then this contribution should be crossing-symmetric by itself, since crossing does not mix terms with half-integer powers of both $u$ and $v$ with terms involving integer powers. More precisely,
\begin{equation}
v^{\Delta_{0}} \left. {\cal G}^{(1)}(u,v) \right|_{\mathrm{st},\mathrm{enh-sing}}= u^{\Delta_{0}} \left. {\cal G}^{(1)}(v,u) \right|_{\mathrm{st},\mathrm{enh-sing}}\,,
\end{equation}
where at leading order $\Delta_{0}$ takes its classical value, which is always integer for the theories under consideration. The small $u,v$ behaviour of the TCB
\begin{equation}
H^{(m)}_1(u,v) \sim u^{1/2} v^{m-1/2}
\end{equation}
implies that the sum over $m$ must run over $m=-\Delta_0+1,-\Delta_0+2,\cdots$. We then propose
\begin{equation}
 \left. {\cal G}^{(1)}(u,v) \right|_{\mathrm{st}}= \kappa H^{(-\Delta_0+1)}_1(u,v)+ \sum_{m=1} \alpha_m H^{(-\Delta_0+1+m)}_1(u,v)\,.
\end{equation}
Quite remarkably, it is possible to satisfy crossing symmetry, as a series expansion in small $u,v$, for a specific choice of the appropriate coefficients $\alpha_m$. For instance, to leading order we find
\begin{equation}
a^{\mathrm{st},(0)}_{s} = \frac{ \kappa }{N}J^{2\Delta_0-2}(1-\frac{1}{12} (\Delta_0 -1) \left(4 \Delta_0^2-14 \Delta_0 +9\right) \frac{1}{J^2}+\cdots) \,.
\end{equation}
For any positive integer $\Delta_0$ the series \eqref{largejexp} truncates, and our result for the OPE coefficients is a polynomial in $J^2$. For general $\Delta_0$ the series can be resummed into
\begin{equation}
\label{opest}
a^{\mathrm{st},(0)}_{s} = \frac{1}{N}  \kappa \frac{s\,\Gamma \left(s+\Delta_0-1\right)}{\Gamma \left(s-\Delta_0+2\right)}\,.
\end{equation}
Note that the overall constant $\kappa$ can be fixed by the known OPE coefficient for the single-trace operator of spin two, which corresponds to the stress tensor. In our derivation we have assumed that odd powers of $1/J$ are not present in the large $J$ expansion of $a^{\mathrm{st},(0)}_{s}$. This is actually a consequence of crossing symmetry and the structure of enhanced singularities. An odd power of $1/J$ would generate an enhanced singularity proportional to $\log^2v$, but under crossing this maps to a piece proportional to $\log^2u$, which cannot be matched with anything at this order in $1/N$ (recall that when we expand in the anomalous dimension as in \eqref{largenexp}, the terms of $n$'th order in the anomalous dimension come with $\log^n(u)$, and start appearing at order $N^{-n}$).

Two important comments regarding (\ref{opest}) are in order. First, this result has been obtained as a resummation around large spin. In principle, it can receive corrections for finite spin. We will return to this momentarily. Second, note that, in the same spirit as in \cite{Maldacena:2011jn,Maldacena:2012sf}, we have proven that symmetries are enough to fix $\langle J_0 J_0 J_s \rangle$, but using only mildly the approximate higher-spin symmetry. However, our analysis does not fix $\Delta_0$; it is valid for all values of the dimension of some single-trace scalar operator. We know independently that approximate higher-spin symmetry can only arise for $\Delta_0=1+O(1/N)$ or  $\Delta_0=2+O(1/N)$ \cite{Maldacena:2011jn,Maldacena:2012sf}.

Having computed the CFT data as above, one can compute the full contribution from single-trace operators, not only its enhanced-singular part. For the two cases of interest we find
\begin{align}
\label{freet}
 \left. {\cal G}^{(1)}(u,v) \right|_{\mathrm{st}} &=\kappa \frac{\pi^2}{4} \left( \sqrt{u} + \sqrt{\frac{u}{v}} \right)\,,\quad\Delta_0=1+\cdots\,,\\
  \left. {\cal G}^{(1)}(u,v) \right|_{\mathrm{st}} &=\kappa \frac{\pi^2}{16} \left(u^{3/2} \left(-\frac{1}{v^{3/2}}-1\right)+\sqrt{u} \left(\frac{1}{v^{3/2}}+v-\frac{1}{\sqrt{v}}-1\right) \right)\,,\quad\Delta_0=2+\cdots \,, \nonumber
\end{align}
where $\kappa$ is not fixed by crossing alone, but can be fixed from the known contribution of the stress-energy tensor to the OPE.\footnote{In particular, $\kappa/N = \frac{16}{\pi^2c_T}$, where $c_T=\tilde N  = \frac{2N\sin(\pi\lambda)}{\pi\lambda}$ in both theories.} Note that through crossing the single-trace operators in the dual channel will contribute to corrections to the OPE coefficients of double-trace operators. These corrections can be computed as above, allowing for a full reconstruction of the four-point correlator, modulo truncated solutions. Furthermore, note that there is no contribution proportional to $\log u$ in \eqref{freet}, so that `on average' double-trace operators do not acquire an anomalous dimension. This conclusion is only true if (\ref{opest}) is extended properly to spin zero. For  $\Delta_0=2$ we find that the OPE coefficient $a_0^{\mathrm{st},(0)}=0$, while for $\Delta_0=1$ this requires $a_0^{\mathrm{st},(0)}=\kappa/2N$ (which is half of the analytically continued value from (\ref{opest})). In the next example we will consider a situation in which the OPE coefficient for the intermediate operator $J_0$ differs from the one implied by (\ref{freet}), so that double-trace operators do  acquire an anomalous dimension to order $1/N$ and \eqref{freet} is modified.
As a final comment, note that our procedure not only allows us to reconstruct the entire correlator, but also gives more directly the CFT data for intermediate operators. 

\subsection{Spectrum of double-trace operators to order $1/N$}
As already mentioned, as we take into account $1/N$ corrections, the twist and OPE coefficients of double-trace operators acquire corrections:
\begin{eqnarray}
\tau_{n,s} &=& 2\Delta_0 + 2n + \frac{1}{N} \gamma^{(1)}_{n,s} + \cdots\,,\\
a_{n,s} &=& a^{(0)}_{n,s} + \frac{1}{N} a^{(1)}_{n,s} + \cdots\,.
\end{eqnarray}
In what follows we will compute the CFT-data at order $1/N$, to all orders in inverse powers of the spin, from the enhanced-singularities, or double discontinuities, of the correlator. In order to see how this works let us consider the conformal block decomposition of the double-trace operators $[J_0 J_0]_{n,s}$ contributing to the correlator
\begin{eqnarray} \label{orderonen}
{\cal G}(u,v) = \sum_{n,s} a^{(0)}_{n,s} f_{\tau^{(0)}_n,s}(u,v) +\frac{1}{N} \left(a^{(1)}_{n,s} f_{\tau^{(0)}_n,s}(u,v) +a^{(0)}_{n,s} \gamma^{(1)}_{n,s} \partial_\tau f_{\tau^{(0)}_n,s}(u,v)  \right) + \cdots\,,
\end{eqnarray}
where $\tau^{(0)}_n=2\Delta_0+2n$. We will now assume that the CFT data can be expanded in inverse powers of $J^2$. This implies that \eqref{orderonen}
can be rewritten as
\begin{eqnarray}\label{uexpansion}
{\cal G}(u,v) = \sum_{m,n}\left( u_{m,n}^{(0)} +u_{m,n}^{(1)} \partial_\tau+ \frac{1}{2}u_{m,n}^{(2)}\partial^2_\tau+\cdots  \right) H_{\tau^{(0)}_n}^{(m)}(u,v)\,,
\end{eqnarray}
where $u_{m,n}^{(p)}$ are the coefficients in the large $J$ expansions of a function
\begin{eqnarray}
U^{(p)}_{n,\bar h} = \sum_m \frac{u_{m,n}^{(p)}}{J^{2m}}\,,
\end{eqnarray}
with ${\bar h} = s + \tau/2$ (or $J^2 = {\bar h} ({\bar h}-1)$). The CFT-data can then be written in terms of the functions  $U^{(p)}_{n,\bar h}$ as
\begin{eqnarray}
a_{n,s} = U^{(0)}_{n,\bar h} + \frac{1}{2}\partial_{\bar h} U^{(1)}_{n,\bar h} + \cdots ,~~~a_{n,s} \gamma_{n,s}= U^{(1)}_{n,\bar h} + \frac{1}{2}\partial_{\bar h} U^{(2)}_{n,\bar h} + \cdots\,,
\end{eqnarray}
or more generally 
\begin{eqnarray}
a_{n,s} (\gamma_{n,s})^p= U^{(p)}_{n,\bar h} + \frac{1}{2}\partial_{\bar h} U^{(p+1)}_{n,\bar h}  + \frac{1}{8}\partial^2_{\bar h} U^{(p+2)}_{n,\bar h} +\cdots\,,
\end{eqnarray}
where $\gamma_{n,s} = \gamma_{n,s}^{(1)}/N + \gamma_{n,s}^{(2)}/N^2+\cdots$.
In a $1/N$ expansion, $U^{(p)}_{n,\bar h} \sim N^{-p}$. We now show how to reconstruct the functions  $U^{(p)}_{n,\bar h}$ from the double discontinuities (or enhanced singularities) of the correlator.

Let us start by analysing the problem at leading order in $1/N$. The enhanced-singularity arises from the identity operator in the dual channel. The equation \eqref{uexpansion} we need to solve is
\begin{eqnarray}\label{leadingone}
\sum_{m,n} u_{m,n}^{(0)}  H_{\tau^{(0)}_n}^{(m)}(u,v) = \frac{u^{\Delta_0}}{v^{\Delta_0}}\,,
\end{eqnarray}
where on both sides of the equation only the enhanced singularities are kept. Having computed the functions $H_{\tau^{(0)}_n}^{(m)}(u,v)$ we can fix the coefficients $u_{m,n}^{(0)}$ by expanding both sides for small $u,v$ and matching enhanced singularities. For instance, for $n=0$ we obtain
\begin{eqnarray}\label{leadingtwo}
U^{(0)}_{0,\bar h}=\frac{4^{\Delta_0 } (2 \bar h-1) \Gamma \left(\Delta_0 +\frac{1}{2}\right) \Gamma (\bar h+\Delta_0 -1)}{\sqrt{\pi } \Gamma (\Delta_0 )^3 \Gamma (\bar h-\Delta_0 +1)}\,. 
\end{eqnarray}
As already mentioned, the CFT data can be equivalently given by an inversion formula \cite{Caron-Huot:2017vep}. This inversion formula is more simply written in terms of cross-ratios $z,\bar z$ such that $z \bar z=u,(1-z)(1-\bar z)=v$, and in appendix \ref{appinv} we describe how to use it in a perturbative setting. For the case of the leading twist double-trace operator this is given by
\begin{eqnarray}
U^{(0)}_{0,\bar h}=\frac{r_{\bar h}}{r_{\frac{\tau}{2}}}\frac{2\bar h-1}{\pi^2} \int_0^1 \frac{d\bar z}{\bar z^2}\,  k_{\bar h}(\bar z)  \dDisc\left[ \left(\frac{\bar z}{1-\bar z} \right)^{\Delta_0}\right]\,,
\end{eqnarray}
where $k_{\bar h}(\bar z) =\bar z^{\bar h} \,_2F_1(\bar h,\bar h,2\bar h;\bar z) $, and the double discontinuity is given by
\begin{eqnarray}
\dDisc\left[ \left(\frac{\bar z}{1-\bar z} \right)^{\Delta_0}\right] =  \left(\frac{\bar z}{1-\bar z} \right)^{\Delta_0} 2 \sin^2(\pi \Delta_0)\,.
\end{eqnarray}
Using the integral representation of the hypergeometric function and performing the integrals, we obtain $U^{(0)}_{0,\bar h}$, which exactly agrees with what we have obtained before. For later convenience, we quote the result for $\Delta_0=1$,
\begin{eqnarray}
U^{(0)}_{0,\bar h}=2(2\bar h-1)\,.
\end{eqnarray}
This gives $a_{0,s}^{(0)}= 2 (2s+1)$. However, note that in general $\Delta_0$ receives corrections in powers of $1/N$ (independently of the corrections $\gamma_{n,s}$ to the double-trace anomalous dimensions). 

\subsubsection{Scalar exchange}

As we have seen in section \ref{stope}, an infinite exchange of single-trace operators with specific OPE coefficients does not generate an (average) anomalous dimension for the double-trace operators. Let us now consider how this is modified when we change the OPE coefficient of the scalar single-trace operator $J_0$. Crossing symmetry \eqref{crossing} implies the following structure of enhanced singularities up to order $1/N$,
\begin{equation}\label{leadingenh}
{\cal G}(u,v)  = \frac{u^{\Delta_0}}{v^{\Delta_0}}+ \frac{1}{N} a_0 \left(\frac{u}{v}\right)^{\Delta_0}  f_{\Delta_0,0}(v,u)+ \cdots\,
\end{equation}
(other channels do not have enhanced singularities at this order).
Note that each term in \eqref{leadingenh} itself will contain a non trivial $N-$dependence, through the $1/N$ expansion of $\Delta_0$ and of $a_0$ (which is related to $\langle J_0 J_0J_0 \rangle^2$), but it is convenient to split the $N$ dependence in this way. The first term (from the identity operator) was taken into account in the previous section. The conformal block for a scalar operator (of any dimension) in a general space-time dimension was given in \cite{Dolan:2000ut}, and in our conventions is equal to
\begin{equation}\label{fdeltazero}
 f_{\Delta,0}(u,v) = r^2_{\frac{\Delta}{2}} u^{\Delta/2}  \sum_{m,n=0} \frac{(\Delta/2)^2_m(\Delta/2)^2_{m+n}}{m! n! (\Delta+1-d/2)_m(\Delta)_{2m+1}} u^m (1-v)^n \,.
\end{equation}
As above in \eqref{leadingone},\eqref{leadingtwo}, we can compute the contribution from the $J_0$ operator in the dual channel to the functions $U^{(0)}_{0,\bar h}$ and $U^{(1)}_{0,\bar h}$. This computation is hard for general $\Delta_0$ but can be done for every integer $\Delta_0$. We will report here the results for the relevant cases of $\Delta_0=1+O(1/N)$ and $\Delta_0=2+O(1/N)$. 

\subsubsection*{CS-scalar theory}

For $\Delta_0=1+O(1/N)$ we obtain from \eqref{leadingenh} and \eqref{fdeltazero}
\begin{equation}
\left. {\cal G}(u,v)  \right|_{enh-sing}= \frac{u^{\Delta_0}}{v^{\Delta_0}}- \frac{1}{N} a_0 \pi \left(\frac{\log u - 4 \log 2-\log(1-v)}{\sqrt{v(1-v)}}u+ \cdots \right)+\cdots \,,
\end{equation}
where in the second term we have assumed $\Delta_0=1$, since the OPE coefficient is already of order $1/N$. Terms with higher powers of $u$ in the parentheses are not shown, as they are very cumbersome. The explicit value of $a_0$ in terms of $\lambda$ and $\lambda_6$ can be extracted from $\langle J_0 J_0 J_0 \rangle$, which was calculated in \cite{Aharony:2012nh}. In our conventions it is given by
\begin{align}
\frac{a_0}{N} = \frac{8}{\tilde{N}\pi^2}\left( \frac{(1+\tilde{\lambda}^2 + a_3)^2}{(1+\tilde{\lambda}^2)^3}-1\right) \,,\quad a_3\equiv \frac{1}{4}\tilde{\lambda}^2(1+\tilde{\lambda}^2)\left(3-\frac{\lambda_6}{8\pi^2\lambda^2}\right) \,, \label{a0}
\end{align}
where $\tilde{N}\equiv \frac{2N\sin(\pi\lambda)}{\pi\lambda}$ and $\tilde{\lambda}\equiv \tan(\frac{\pi\lambda}{2})$.\footnote{The full contribution to the four-point function due to $J_0$ exchange is $a_0^{\mathrm{full}} = a_0^{\mathrm{st}} + a_0$, where $a_0^{\mathrm{st}}=\kappa/2$ is given by half the naive value in \eqref{opest}, and $a_0$ in \eqref{a0}. The $a_0^{\mathrm{st}}$ piece combines with contributions of the other single-trace currents to reproduce the part of the four-point function proportional to the one of the free theory.}

Surprisingly, the anomalous dimension as well as the correction to the OPE coefficients vanish for all double-trace operators with $n>0$. The corrections for the leading twist double-trace operators can again be computed in two equivalent ways: expanding the CFT data in powers of $1/J$ and using large spin perturbation theory to fix it order by order, or with the help of the inversion formula. In appendix \ref{appinv} we describe the precise inversion integrals to be computed in general. For the present case,  the hardest element of the computation involves the integral
\begin{equation}
\frac{r_{\bar h}(2\bar h-1)}{\pi^2} \int_0^1 \frac{d\bar z}{\bar z^2}\,  k_{\bar h}(\bar z) \dDisc\left[ \frac{\sqrt{\bar z} \log \bar z}{\sqrt{1-\bar z}} \right] = \frac{8 \left(\frac{1}{2 \bar h-1}-\psi ^{(0)}(\bar h)+\psi ^{(0)}\left(\bar h-\frac{1}{2}\right)\right)}{\pi }\,,
\end{equation}
which can be explicitly confirmed by expanding $\frac{\sqrt{\bar z} \log \bar z}{\sqrt{1-\bar z}}$ in powers of $\zeta= \frac{1-\bar z}{\bar z}$. Each power of $\zeta$ can be integrated by using the integral representation of the hypergeometric function, and the resulting sum can be seen to have the same large $\bar h$ expansion as the right-hand side. Alternatively, one can also compute both sides to a high numerical precision, for finite values of $\bar h$. 

With this result at hand, we can compute the relevant CFT data. For $U^{(0)}_{0,\bar h}$ at order $1/N$, the contribution due to exchange of $J_0$ is found to be
\begin{equation}
U^{(0)}_{0,\bar h}= \frac{8 a_0}{N} \left(\frac{1}{2 \bar h-1}-\Psi(\bar{h}) -1 \right)\,,\label{U0exchange}
\end{equation}
where we have introduced
\begin{eqnarray}
\Psi(\bar{h})=\psi ^{(0)}(\bar h) -\psi ^{(0)}\left(\bar h-\frac{1}{2}\right) - \log (4)\,.\label{Psi}
\end{eqnarray}
The function $\Psi(\bar{h})$ will also appear in the next section.\footnote{We emphasize that $U^{(0)}_{0,\bar{h}}$ also receives other $1/N$ corrections due to the (unknown) anomalous dimension ($\Delta_0-1$) of $J_0$ (see, e.g., \eqref{leadingtwo}). The above computations only capture the contributions to anomalous dimensions and OPE coefficients that are proportional to $a_0$, which arise due to the exchange of $J_0$ in the OPE.} For $U^{(1)}_{0,\bar h}$ at order $1/N$ we obtain
\begin{equation}
U^{(1)}_{0,\bar h} = -\frac{8a_0}{N}\,.
\end{equation}
At order $1/N$ we also find $U^{(0)}_{n,\bar h}=U^{(1)}_{n,\bar h}=0$ for $n>0$. As above, this implies that only the leading twist ($n=0$) double-trace operators acquire anomalous dimensions and corrections to their OPE coefficients due to the exchange. Moreover, because we found that $U^{(1)}_{0,\bar{h}}$ is independent of $\bar{h}$ at order $1/N$, the corrections to the OPE coefficients at this order are fully captured by $U^{(0)}_{0,\bar{h}}$:
\begin{align}
a^{(1)}_{0,s} = U^{(0)}_{0,\bar{h}=s+1} = \frac{8a_0}{N}\left(\frac{1}{2s+1} + \psi^{(0)}\left(s+\frac{1}{2}\right)-\psi^{(0)}(s+1)+\log(4)-1\right) \,,
\end{align}
where we have used that $\bar h=s+1$ at leading order in $1/N$. 

From the expression for $U^{(1)}_{0,\bar h}$ to order $1/N$ and  from $U^{(0)}_{0,\bar h}$ to leading order, we can compute the anomalous dimensions for double-trace operators to leading order due to the exchange. We obtain
\begin{equation}
\gamma_{0,s}^{(1)} = -\frac{4 a_0}{2s+1}\,.\label{gammaexchange}
\end{equation}
The result \eqref{gammaexchange} already appears in \cite{Giombi:2017rhm} (with $a_0$ in \eqref{a0} evaluated at $a_3=0$), though here we saw that in addition $\gamma_{n>0,s}^{(1)}=0$. Alternatively, the same result for $\gamma_{0,s}^{(1)}$ can be obtained by decomposing in conformal blocks the four-point function given explicitly in terms of $\bar{D}$-functions in \cite{Turiaci:2018nua}. Moreover, from that decomposition it is also possible to verify that $\gamma_{n>0,s}^{(1)}$ vanishes.

\subsubsection*{General case}
Exactly the same steps can be repeated for other cases of integer $\Delta_0$. In this case $U^{(0)}_{n,\bar h}$  and $U^{(1)}_{n,\bar h}$ do receive contributions also for $n>0$. We quote the results for $\Delta_0=2$, $n=0$ and to leading order in $1/N$:
\begin{eqnarray}
U^{(0)}_{0,\bar h} &=&  -\frac{20 a_0}{N} (2\bar h-1)\,,\\
U^{(1)}_{0,\bar h} &=&  -\frac{24 a_0}{N} (2\bar h-1)\,.
\end{eqnarray}
From the result \eqref{leadingtwo} for $U^{(0)}_{n,\bar h}$ to zeroth order we can then compute the anomalous dimension due to the exchange of $J_0$. We obtain
\begin{equation}
\gamma_{0,s}^{(1)} = -\frac{2 a_0}{{\bar h}({\bar h}+1)} = -\frac{2 a_0}{(s+1)(s+2)}\,.
\end{equation}

\bigskip

\noindent  As for the contributions due to single-trace operators, it is again possible to resum the CFT-data in order to reconstruct the whole correlator. In addition one will always have the freedom to add solutions consistent with crossing which do not produce enhanced singularities in either channel. These correspond to solutions where only operators up to a finite spin $L$ receive corrections, and can be easily constructed following the algorithm in \cite{Heemskerk:2009pn}. From the point of view of crossing symmetry there is no clear way to constraint $L$. On the other hand, consistency with the Regge limit forbids all solutions with $L>2$, see \cite{Caron-Huot:2017vep} and a related discussion in \cite{Turiaci:2018nua}.

\section{Constraints on the spectrum to order $1/N^2$}
In this section we work out the consequences of crossing symmetry for the spectrum of double-trace operators to order $1/N^2$. The standard lore is that the CFT data for single-trace operators should fix completely the form of the correlator (up to low-spin truncated solutions). However, this CFT data may need to include information from more general single-trace correlators. In this section we will study the constraints that come just from the specific correlator $\langle J_0 J_0 J_0 J_0 \rangle$. We will start by drawing general conclusions for the CS-matter theories. Later we will focus on the scalar case, with $\lambda=0$, for which we can obtain sharper results. 

\subsection{CS-matter theories}
Let us start by considering the case in which the CS coupling constant $\lambda$ is different from zero. In this case single-trace operators acquire an anomalous dimension of order $1/N$. This has been computed in \cite{Giombi:2016zwa}. For even spin $s$ the result at order $1/N$ takes the form
\begin{equation}\label{stanomdim}
\gamma_s^{\mathrm{st},(1)} =  \alpha(\lambda) \alpha_s+\beta(\lambda)   \beta_s\,,
\end{equation}
with
\begin{equation}
\alpha(\lambda)=\frac{\pi \lambda}{4} \tan\left(\frac{\pi \lambda}{2} \right),~~~\beta(\lambda)=\frac{\pi \lambda}{8}\sin(\pi \lambda)\,,
\end{equation}
and
\begin{eqnarray}
\alpha_s&=& \frac{16 (s-2)}{3 \pi ^2 (2 s-1)},\\
\beta_s &=& \frac{2}{\pi^2} \sum_{n=1}^s \frac{1}{n-\frac{1}{2}}+\left(\frac{2}{3 \pi^2}\right)\frac{-38 s^4+24 s^3+34 s^2-24 s-32}{\left(s^2-1\right) \left(4 s^2-1\right)}\,.
\end{eqnarray}
The result is the same for both Chern-Simons-matter theories. In addition, the OPE coefficients of single-trace operators are also expected to acquire corrections. In the present paper we will not attempt to rederive the above results. Rather, we will limit ourselves to studying their consequences for the spectrum and OPE coefficients of higher twist operators. The anomalous dimension \eqref{stanomdim} admits a large $J$ expansion, with even and odd powers of $1/J$ and also a logarithmic term at leading order:
\begin{eqnarray}\label{gammaone}
\gamma_s^{\mathrm{st},(1)} = \frac{\beta(\lambda)}{\pi^2} \log J^2 + \rho + 4\frac{\beta(\lambda)-\alpha(\lambda)}{\pi^2} \frac{1}{J}+ \cdots\,,
\end{eqnarray}
where $\rho$ is some calculable constant and $J^2=s^2-1/4$. A similar behaviour is true for $a^{\mathrm{st},(0)}_{s} \gamma_s^{\mathrm{st},(1)}$. We assume that the correction $a_s^{\mathrm{st},(1)}$ to the OPE coefficients admits a similar expansion. Hence, it follows from the conformal partial wave decomposition, that the contribution to order $1/N^2$ arising from the single-trace operators can be written in terms of twist conformal blocks. For the CS-Scalar theory \eqref{gammaone} gives
\begin{eqnarray}
 \left. {\cal G}^{(2)}(u,v) \right|_{\mathrm{st}} = \frac{\beta(\lambda)}{\pi^2} \kappa \partial_\tau H_1^{(0,\log J^2)}(u,v) + \kappa \rho \partial_\tau H_1^{(0)}(u,v)+ 4 \kappa \frac{\beta(\lambda)-\alpha(\lambda)}{\pi^2} \partial_\tau H_1^{(1/2)}(u,v) + \cdots\,,
\end{eqnarray}
together with terms without the $\tau$-derivative, arising from corrections to the OPE coefficients. We have introduced $H_1^{(0,\log J^2)}(u,v) = -\left. \partial_m H_1^{(m)}(u,v) \right|_{m=0}$, which accounts for the logarithmic terms. We are working here in a normalisation in which the contribution from single-trace operators to leading order is
\begin{eqnarray}
G(u,v)_{\mathrm{st}} = \kappa \frac{\pi^2}{4} \sqrt{\frac{u}{v}} + \cdots,
\end{eqnarray}
as in section \ref{stope}. From the specific form (\ref{stanomdim}) if follows that the large $J$ expansion of $\gamma_s$ contains only even powers of $1/J$ at leading order in $\lambda$, while for generic $\lambda$, and starting at order $\lambda^4$, the expansion contains also odd powers. The consequences of both terms in the crossing equations are quite distinct and can be analysed separately. 

Even powers of $1/J$ produce enhanced singularities of the form
\begin{equation}
H_1^{(m)}(u,v) \sim u^{n_1+1/2}v^{n_2+1/2}
\end{equation}
for integers $m,n_1,n_2$. For integer classical $\Delta_0$, these enhanced singularities can only map to themselves under crossing symmetry. Furthermore, the double-trace operators of approximate even twist cannot generate them. This implies the appearance of new operators with odd twist $\tau=3,5,\cdots$ at order $1/N^2$.  These new operators appear with their classical twist, and crossing symmetry gives an infinite number of constraints on the OPE coefficients with which they appear. More precisely, the OPE coefficients contain a logarithmic piece in a large $J$ expansion, which is completely fixed by crossing symmetry. Exactly the same analysis and conclusions hold for the CS-Fermion theory. Actually, the logarithmic part of the OPE coefficient for odd-twists operators admits universal expressions. For instance, for $\tau=3,5$ we obtain at order $1/N^2$
\begin{eqnarray}
a^{\tau=3}_s &=& a_s^{\mathrm{st},(0)} \frac{\kappa}{\pi^2} \left(32 \alpha(\lambda)+34 \beta(\lambda) + \frac{\beta(\lambda)}{2 J^2} \right) \log J^2  + \cdots\,, \\
a^{\tau=5}_s &=& a_s^{\mathrm{st},(0)} \frac{\kappa (2\Delta_0-1)^2}{9 \pi^2} \left(256 \alpha(\lambda)+2336 \beta(\lambda) - \frac{64 (\alpha(\lambda)+\beta(\lambda))}{2 J^2} \right) \log J^2  + \cdots\,,
\end{eqnarray} 
where $a_s^{\mathrm{st},(0)}$ are the OPE coefficients for the single-trace operators at order $1/N$, and $\Delta_0$ the classical dimension of $J_0$. These expressions are valid for both the CS-Scalar and CS-Fermion theories, and for all values of $\lambda$. Note that they should be interpreted in an average sense, since the double-trace operators with twist three and higher can be degenerate. Similar results are obtained for operators of other odd twists. Furthermore, crossing symmetry also constrains the corrections to the OPE coefficients of the single-trace operators with $\tau=1$, and fixes the logarithmic part, as before. The expression is somewhat complicated, but it depends on $\gamma^{(1)}_{s=0}$, $\alpha(\lambda)$ and $\beta(\lambda)$. Let us mention that in the limit in which the anomalous dimension disappears, crossing symmetry is consistent with the fact that the corrections to the single-trace OPE coefficients also vanish. 

Let us analyse now the presence of odd powers of $1/J$. Those first arise at order $\lambda^4$ in a perturbative expansion. This has important consequences. The enhanced singularity from single-trace operators, due to odd powers of $1/J$ in their anomalous dimension, will be proportional to $\partial_\tau H^{(m=1/2)}_{\tau=1}(u,v)$, $\partial_\tau H^{(m=3/2)}_{\tau=1}(u,v)$, etc. For half-integer $m$, $H^{(m)}_{\tau}(u,v)$ contains terms proportional to $\log^2 v$. Hence, via crossing symmetry, the presence of odd powers of $1/J$ in the anomalous dimensions of single-trace operators (and in particular the presence of $1/J$) implies the following terms in the correlator to order $1/N^2$:
\begin{equation}
G^{(2)}(u,v)  \sim \log^2 (u) u^n \frac{\log v}{v^{1/2}}\,.
\end{equation}
This contribution can only arise from the square of the anomalous dimensions of double-trace operators with even twist. Also, note that the $v$ dependence implies an enhanced singularity as $v \to 0$, which can only arise if double-trace operators with arbitrarily large spin acquire an anomalous dimension. This may seem puzzling in situations where double-trace operators do not acquire an anomalous dimension to order $1/N$ (the CS-Fermion theory, and the CS-Scalar theory for a specific value of $\lambda_6$). The resolution to this puzzle is that double-trace operators are generally degenerate, and there is necessarily more than one species for given quantum numbers. In average, the anomalous dimensions of all operators with the same spin and classical twist may cancel out to leading order (since they don't need to have the same sign), but at the next order, the average of their squares will not vanish. Note that this effect starts at order $\lambda^4$, consistent with an anomalous dimension of order $\lambda^2$.  

\subsection{Scalar $\phi^6$ theory}
Let us now focus on the model \eqref{lambdasix} with $\lambda = 0$ and $\lambda_6 \neq 0$. For $\lambda_6=0$ the correlation function is that of the free scalar theory, in which $G^{(2)}(u,v)$ and all higher orders vanish, but this is no longer true for finite $\lambda_6$. In this theory, because of parity, we do not expect new operators of odd twist to appear at order $1/N^2$, which vastly simplifies the problem. Let us start by discussing the single-trace operators. For  $\lambda = 0$ the single-trace operators do not acquire an anomalous dimension to order $1/N$. In addition, as discussed in section \ref{stope}, crossing symmetry is consistent with the fact that their OPE coefficients do not get corrected to order $1/N$. More precisely, assume that the corrections to the OPE coefficients admit an expansion in powers of $1/J$ (with possible logarithmic terms). The even powers are fully constrained by crossing symmetry, and in the absence of anomalous dimensions they can only be proportional to the leading order answer, which we already computed. Note that such a contribution, even if present, would not produce an anomalous dimension for double-trace operators, since it is there already in the free theory. Odd powers of $1/J$ are less constrained by crossing symmetry, as their contribution maps to double-trace operators. As we will see below, the leading $1/J$ power is again forbidden. It is then reasonable to assume (although we are not proving it) that indeed the correction to the single-trace OPE coefficients vanishes. Note that this is far from obvious from the field theory point of view. Note furthermore that we can interpret our result for $\langle J_0  J_0 J_s \rangle$ as a consequence of the higher spin symmetry. If the tower of conserved currents breaks at order $1/N^2$ rather than order $1/N$, which is the case for $\lambda=0$ where the non-conservation of the currents is proportional to a triple-trace operator, then we expect the corrections to the OPE coefficients to be of the same order, with respect to the leading result. 

Let us assume that this is the case and compute the double-discontinuity of the correlator to order $1/N^2$. As computed in the previous section for the case $\Delta_0=1+\cdots$, and in the absence of truncated solutions, the anomalous dimension is non-vanishing only for the leading twist double-trace operators, of approximate twist two. The double discontinuity would then arise, via crossing, from the coefficient of $\log^2u$ in the $4$-point function, which is
\begin{equation}\label{gtwosum}
\left. {\cal G}^{(2)}(u,v) \right|_{\log^2u}=\sum_{s} a_{0,s}^{(0)} \frac{(\gamma_{0,s}^{(1)})^2}{8} f_{\tau=2,s}(u,v)\,.
\end{equation}
For each value of the spin one should sum here over all the degenerate operators with approximate twist two. In order to compute the sum above, one would thus need to solve a complicated mixing problem, which requires the knowledge of general correlators $\langle J_{s_1} J_{s_2} J_{s_3} J_{s_4} \rangle$ to order $1/N$. However, for large $\lambda_6$ we are able to argue (see appendix \ref{nomixing}) that in the present case, with $\lambda=0$, the operators $[J_0 J_0]_{n,s}$ that acquire an anomalous dimension are non-degenerate (at leading order in $1/\lambda_6$), and hence the square of their anomalous dimension appearing in \eqref{gtwosum} can be easily computed from \eqref{gammaexchange} at this order, and is simply
\begin{equation}
(\gamma_{0,s}^{(1)})^2 = 16 \frac{a_0^2}{(2s+1)^2}+O(\lambda_6^3)\,.
\end{equation}
This still leads to a very complicated sum in \eqref{gtwosum}, but in appendix \ref{appsums} we show how to deal with it. We will be interested in computing the terms of the sum that, after crossing symmetry, contribute to the anomalous dimension of the leading twist double-trace operators. In particular we are interested in the small $v$ limit of the above sum. It is convenient to express the result in terms of the cross-ratios $z,
\bar z$. In the limit of interest we find:
\begin{equation}
\left. {\cal G}^{(2)}(z,\bar z) \right|_{\log^2 z} = h_0(z) \log^2(1-\bar z)+h_1(z) \log (1-\bar z) + \cdots\,.
\end{equation}
The expression for $h_0(z)$ and $h_1(z)$ will suffice to compute the anomalous dimension of the leading twist double-trace operators for large $\lambda_6$ to order $1/N^2$; in all the equations below the prefactor $a_0^2$ is correct  at leading order for large $\lambda_6$ (order $\lambda_6^4$) but not at lower orders where mixing affects the result. By using the strategy outlined in appendix \ref{appsums}  we can compute $h_0(z),h_1(z)$ as a series expansion in powers of $z$, to any desired order. With enough effort we can actually resum the series and obtain
\begin{eqnarray}
h_0(z) &=& 4a_0^2 \frac{z K(z)}{8 \pi  \sqrt{1-z}}\,,\\
h_1(z)&=&4a_0^2 \frac{z \left(-4 K\left(\frac{1}{2}-\frac{\sqrt{1-z}}{2}\right)^2+\pi  K(1-z)- K(z) \left(\log\left(\frac{1-z}{z} \right)+\log (128)\right)\right)}{8 \pi  \sqrt{1-z}}\,, \nonumber
\end{eqnarray}
where $K$ is the complete elliptic integral of the first kind. As explained in the previous section, we can now compute the contribution of the resulting double-discontinuity, after crossing, to the functions $U^{(2)}_{0,\bar h}$ and  $U^{(1)}_{0,\bar h}$. Let us start with $U^{(2)}_{0,\bar h}$. Following the formulae given in appendix \ref{appinv} we see that the relevant inversion integral is
\begin{eqnarray}
\frac{r_{\bar h}(2\bar h-1)}{\pi^2} \int_0^1 \frac{d\bar z}{\bar z^2}\,  k_{\bar h}(\bar z) \sqrt{\bar z} K(1-\bar z) = \frac{2}{\pi}\frac{1}{2\bar h-1}\,
\end{eqnarray}
(the simplest way to check this integral is as an expansion around large $\bar h$, to any desired order). This leads to
\begin{equation}
U^{(2)}_{0,\bar h}= \frac{1}{N^2}\frac{32 a_0^2}{2\bar h-1}\,,
\end{equation}
which exactly reproduces the precise value of $a^{(0)}_{0,s}(\gamma_{0,s}^{(1)})^2$!  The inversion integrals relevant to the computation of $U^{(1)}_{0,\bar h}$ are much harder, but they can be evaluated with some effort. We obtain 
\begin{eqnarray}
& \frac{r_{\bar h}(2\bar h-1)}{\pi^2} \int_0^1 \frac{d\bar z}{\bar z^2}\,  k_{\bar h}(\bar z) \sqrt{\bar z}K^2\left(\frac{1}{2}-\frac{\sqrt{\bar z}}{2}\right)= \frac{\Gamma^2\left(\frac{\bar h}{2}\right)}{4 \Gamma^2\left(\frac{\bar h}{2}+\frac{1}{2}\right)}\,,\\
& \frac{r_{\bar h}(2\bar h-1)}{\pi^2} \int_0^1 \frac{d\bar z}{\bar z^2}\,  k_{\bar h}(\bar z) \sqrt{\bar z}\left( \pi K(\bar z) + K(1-\bar z)\left(\log \left(\frac{1-\bar z}{\bar z}\right) -12 \log 2\right)\right)= \frac{8(2\bar h-1)\Psi(\bar h)-1}{\pi(2\bar h-1)^2}\,,
\end{eqnarray}
where $\Psi(\bar h)$ was defined in \eqref{Psi}. Again, these integrals can be checked in a large $J^2=\bar h (\bar h-1)$ expansion to any desired order, and also numerically with high precision. With these results at hand we can readily compute the contribution at order $1/N^2$ to $U^{(1)}_{0,\bar h}$. We obtain
\begin{equation}
U^{(1)}_{0,\bar h}= \frac{4a_0^2}{N^2} \left(8 \frac{(2\bar h-1) \Psi(\bar h)+(2\bar h-2)}{(2\bar h-1)^2}- \frac{\pi  \Gamma \left(\frac{\bar h}{2}\right)^2}{\Gamma \left(\frac{\bar h+1}{2}\right)^2}\right)\,.
\end{equation}
We are now ready to assemble all the results to obtain the anomalous dimensions of the double-trace operators, of twist 2, to order $1/N^2$. The full result to this order (and at large $\lambda_6$) reads
\begin{equation}
\gamma_{0,s}^{(2)} =- \frac{4 a_0}{N} \frac{1}{2\bar h-1}  + \frac{2 a_0^2}{N^2}\left( \frac{8}{(2\bar h-1)^3} - \frac{\pi  \Gamma \left(\frac{\bar h}{2}\right)^2}{(2\bar h-1)\Gamma \left(\frac{\bar h+1}{2}\right)^2} \right) + \cdots\,.
\end{equation}
This result looks even nicer simply in terms of the spin $s$. Recalling $\bar h = s+1 - \frac{2 a_0}{N} \frac{1}{2s+1} +\cdots $ we obtain
\begin{equation}
\gamma_{0,s} =- \frac{4 a_0}{N} \frac{1}{2s+1}  - \frac{2 a_0^2}{N^2}\frac{\pi  \Gamma \left(\frac{s+1}{2}\right)^2}{(2s+1)\Gamma \left(\frac{s+2}{2}\right)^2} + \cdots\,,
\end{equation}
where for the Scalar theory $a_0=\frac{\lambda_6(\lambda_6-256)}{2^{12}}+O(1/N)$.

\section*{Acknowledgements}

We would like to thank C.~Sleight and M.~Taronna for useful discussions and for correcting a mistaken assumption in the first version of this paper.
RY would like to thank A. Zhiboedov for useful discussions, and the Annual Meeting of the Simons Collaboration on the Nonperturbative bootstrap for hospitality.
The work of OA and RY was supported in part  by the I-CORE program of the Planning and Budgeting Committee and the Israel Science Foundation (grant number 1937/12), by an Israel Science Foundation center for excellence grant, and by the Minerva foundation with funding from the Federal German Ministry for Education and Research. OA is the Samuel Sebba Professorial Chair of Pure and Applied Physics. The work of AB was supported in part by the Knut and Alice Wallenberg Foundation under grant KAW 2016.0129. LFA would like to thank Uppsala University for hospitality during part of this work. AB would like to thank the University of Oxford and the Weizmann Institute of Science for hospitality while parts of this work have been done. 

\appendix

\section{Twist conformal blocks}
\label{appTCB}

In this appendix we consider the sums over conformal blocks
\begin{equation}
H_\tau(u,v) \equiv \sum_{s=0,2,\cdots} f_{\tau,s}(u,v)\,.
\end{equation}
We find it convenient to use conventions in which 
\begin{equation}
f_{\tau,s}(u,v) = r_{\frac{\tau}{2}+s}r_{\frac{\tau}{2}} \tilde f_{\tau,s}(u,v),~~~r_h=\frac{\Gamma(h)^2}{\Gamma(2h)}\,,
\end{equation}
where $\tilde f_{\tau,s}(u,v)$ has the usual normalisation, namely $\tilde f_{\tau,s}(u,v)  \to u^{\tau/2}$ as $u \to 0$ and $v \to 1$. We also consider their generalisations 
\begin{equation}
H^{(m)}_\tau(u,v) \equiv \sum_{s=0,2,\cdots} \frac{ f_{\tau,s}(u,v)}{J^{2m}}\,,
\end{equation}
where the conformal spin is defined as $J^2=(s+\tau/2)(s+\tau/2-1)$. In this appendix we will only focus in their enhanced-singular part, which has a double-discontinuity, and all equalities are to be understood up to regular terms ({\it i.e.} terms which do not lead to a double-discontinuity). These sums satisfy several properties which we now list, and which allow the computation of their enhanced-singular part. Let us start with $H_\tau(u,v)$. Its small $u$ behaviour can be computed from the explicit form of collinear-conformal blocks, which are independent of the number of dimensions. We obtain
\begin{equation}
H_\tau(u,v) = u^{\frac{\tau}{2}}\frac{\pi  \Gamma \left(\frac{\tau }{2}\right)^2 (1-v)^{\frac{1}{2}-\frac{\tau }{2}}}{4 \sqrt{v} \Gamma (\tau )} +\cdots\,.
\end{equation}
Note that this enhanced-singular contribution admits an expansion in half-integer powers of $v$. This is true for higher powers of $u$ as well. Hence, $H_\tau(u,v)$ admits an expansion of the form
\begin{equation}
\label{expansion}
H_\tau(u,v) = u^{\frac{\tau}{2}}\frac{\pi  \Gamma \left(\frac{\tau }{2}\right)^2 (1-v)^{\frac{1}{2}-\frac{\tau }{2}}}{4 \sqrt{v} \Gamma (\tau )}\left(1+ \sum_{p=1,q=0}^{\infty} a_{p,q}u^p v^q \right)\,.
\end{equation}
Second, the function $H_\tau(u,v)$ is the eigenfunction of a quartic operator constructed in \cite{Alday:2016njk}:
\begin{equation}
\label{quartic}
{\cal H}_\tau H_\tau(u,v) = \lambda_\tau H_\tau(u,v) ,
\end{equation}
where  
\begin{equation}
{\cal H}_\tau ={\cal D}_4-{\cal D}_2^2+(d^2-d (2 \tau +3)+\tau ^2+2 \tau +2){\cal D}_2\,.
\end{equation}
Here ${\cal D}_4,{\cal D}_2$ are standard quartic and quadratic Casimirs, see for instance \cite{Hogervorst:2013kva}, and are given (in the variables $z$,$\bar z$ defined by $u=z{\bar z}$, $v=(1-z)(1-{\bar z})$) by
\begin{eqnarray}
{\cal D}_2 = D+ \bar D+(d-2)\frac{z \bar z}{z-\bar z}((1-z)\partial-(1-\bar z){\bar \partial})\,,\\
{\cal D}_4 =\left(\frac{z \bar z}{z-\bar z} \right)^{d-2} (D-\bar D) \left(\frac{z \bar z}{z-\bar z} \right)^{2-d}(D-\bar D)\,,
\end{eqnarray}
with $D=(1-z)z^2\partial^2-z^2\partial$, and $\partial = \partial /\partial z$. The eigenvalue is
\begin{equation}
\lambda_{\tau} =\frac{1}{4} \tau  (\tau +2-2d) (\tau -d) (\tau +2-d)\,.
\end{equation}
Plugging (\ref{expansion}) into (\ref{quartic}) we can solve for the coefficients $a_{p,q}$ to any desired order. Let us now turn to the functions $H^{(m)}_\tau(u,v)$. For $m=0$ they reduce to $H_\tau(u,v)$. Furthermore, they are related by the recurrence relation 
\begin{equation}
\label{recursion}
{\cal C}_{\tau} H^{(m+1)}_\tau(u,v) =H^{(m)}_\tau(u,v)\,,
\end{equation}
where the Casimir operator ${\cal C}_{\tau}$ is given by
\begin{equation}
{\cal C}_{\tau} = {\cal D}_2+\frac{1}{4}\tau(2d-\tau-2)\,.
\end{equation}
The form of $H_\tau(u,v)$ suggests the following expansion
\begin{equation}
H^{(m)}_\tau(u,v) =\frac{v^{m-\frac{1}{2}} \Gamma \left(\frac{1}{2}-m\right)^2 \Gamma \left(\frac{\tau }{2}\right)^2 u^{\tau /2} (1-v)^{\frac{1}{2}-\frac{\tau }{2}}}{4 \Gamma (\tau )}\sum_{p,q=0} a_{p,q}^{(m)} u^p v^q\,.
\end{equation}
Plugging this expansion into the recursion relation (\ref{recursion}) and using the $m=0$ case as an initial condition, one can recursively find all coefficients $a_{p,q}^{(m)}$. Note that this expression is analytic in $m$ and valid for general $\tau$. 

\section{Inversion formula}
\label{appinv}
In this appendix we review the basics of the inversion formula derived in \cite{Caron-Huot:2017vep}, and its reduction to the case of leading twist operators in a perturbative expansion. It is convenient to work in terms of cross-ratios $(z,\bar z)$ such that $u= z \bar z$ and $v=(1-z)(1-\bar z)$. We start by writing down the inversion formula for the case of identical operators:
\begin{equation}
c(s,\Delta) = \frac{2\bar h-1}{\pi^2}\frac{r_{\bar h}}{r_{h} } \int_0^1 \frac{dz}{z^2} \frac{d\bar z}{\bar z^2} \left| \frac{z-\bar z}{z\bar z} \right|^{d-2} f_{-2 + 2 d + s - \Delta,1 - d + \Delta}(z,\bar z) \dDisc\left[{\cal G}(z,\bar z) \right]\,,
\end{equation}
where in this appendix $f_{\tau,s}(z,\bar z)$ are the usual conformal blocks with the standard normalisation, and we have introduced variables $h=\frac{\Delta-s}{2},\bar h=\frac{\Delta+s}{2}$. The idea is that the function $c(s,\Delta) $ has poles in the $\Delta-$plane, located at the spectrum of the intermediate operators, and the residue at these poles corresponds to the corresponding OPE coefficient squared
\begin{equation}
c(s,\Delta) \sim \frac{a_{\Delta^*}}{\Delta-\Delta^*}\,.
\end{equation}
We will be interested in using the inversion formula for the case of leading twist double-trace operators. In a perturbative setting they have twist $2\Delta_0+\gamma_s$. Their contribution can be singled out by considering the small $z$ limit in the integrand above. In this limit the conformal block reduces to its collinear part
\begin{equation}
f_{-2 + 2 d + s - \Delta,1 - d + \Delta} (z,\bar z)= z^{\frac{-2 + 2 d + s - \Delta}{2}} k_{\frac{s+\Delta}{2}}(\bar z) + \cdots\,,
\end{equation}
and the correlator has an expansion of the form
\begin{equation}
{\cal G}(z,\bar z) = z^{\Delta_0}\left(h_0(\bar z) + \log (z) h_1(\bar z)+\log^2 (z) h_2(\bar z)  +\cdots \right)\,.
\end{equation}
Plugging these two expansions into the inversion formula and performing the integrals over $z$ we obtain
\begin{eqnarray}
c(s,\Delta) &=& -\frac{1}{h-\Delta_0} \frac{2\bar h-1}{\pi^2}\frac{r_{\bar h}}{r_{h} }\int_0^1 \frac{d\bar z}{\bar z^2}k_{\bar h}(\bar z) \dDisc\left[h_0(\bar z) \right]  \nonumber\\ \label{invpert}
& &-  \frac{1}{(h-\Delta_0)^2} \frac{2\bar h-1}{\pi^2}\frac{r_{\bar h}}{r_{h} } \int_0^1 \frac{d\bar z}{\bar z^2}k_{\bar h}(\bar z) \dDisc\left[h_1(\bar z) \right]\\
& &-  \frac{2}{(h-\Delta_0)^3} \frac{2\bar h-1}{\pi^2}\frac{r_{\bar h}}{r_{h} } \int_0^1 \frac{d\bar z}{\bar z^2}k_{\bar h}(\bar z) \dDisc\left[h_2(\bar z) \right]+\cdots \nonumber\,.
\end{eqnarray}
The appearance of higher order poles in a perturbative setting exactly agrees with our expectations. Indeed
\begin{equation}
c(s,\Delta) \sim \frac{a_s}{\Delta-(2\Delta_0+s+\gamma_s)} =  \frac{a_s}{\Delta-(2\Delta_0+s)} +  \frac{a_s \gamma_s}{(\Delta-(2\Delta_0+s))^2}+\cdots\,.
\end{equation}
Since the integrand over $\bar z$ in the inversion formula depends on $(\Delta,s)$ only through $\bar h$, it is convenient to expand $c(s,\Delta)$ around its poles keeping $\bar h$ fixed. More precisely, expanding $c(s,\Delta)$ around the pole at $h$ for fixed $\bar h$, we obtain
\begin{equation}
c(s,\Delta) =-\frac{U^{(0)}_{0,\bar h}}{h-\Delta_0} - \frac{U^{(1)}_{0,\bar h}}{2(h-\Delta_0)^2} - \frac{U^{(2)}_{0,\bar h}}{4(h-\Delta_0)^3} + \cdots \,.
\end{equation}
Comparing this expansion to (\ref{invpert}) we obtain integral expressions for $U^{(0)}_{0,\bar h},U^{(1)}_{0,\bar h},\cdots$ in terms of the double-discontinuities of the correlator. 

\section{No mixing in $\phi^6$ theory at large $\lambda_6$}
\label{nomixing}
In this appendix we discuss the mixing of double-trace operators in the $\phi^6$ theory at large $N$. At leading order in $1/N$ the four-point correlator $\langle J_0J_0J_0J_0 \rangle$ is simply that of generalized free fields. Then, at order $1/N$, we have the exchange of $J_0$, which (as discussed in the main text) produces an anomalous dimension for the double-trace operators of twist 2 (only!), proportional to 
\begin{equation}
\gamma_{0,s}^{(1)} \sim \frac{1}{2s+1}\,.
\end{equation}
In general this is an average anomalous dimension over all operators that mix with $[J_0 J_0]_{0,s}$, and we would like to understand which operators contribute.

The double-trace operators of twist two take the general form
\begin{equation}
[J_{s_1} J_{s_2}]_\ell \sim J_{s_1} \partial^\ell J_{s_2}\,,
\end{equation}
where the derivatives act in both directions (so as to produce a primary operator), and no indices are contracted (because we are looking at twist two operators, of spin $s_1+s_2+\ell$). For an even spin $s$, the operators we consider are
\begin{equation}
[J_0 J_0]_s,[J_0J_2]_{s-2}, \cdots, [J_0 J_s]_0, \cdots\,
\end{equation}
where the dots at the end denote operators of the form $[J_{s_1}J_{s_2}]_\ell$ with both $s_1,s_2>0$. In the large $N$ limit these operators are orthogonal to each other, since the currents themselves are. We further assume a normalization where they are orthonormal. The eigenfunctions of the scaling operator are linear combinations of those operators, with anomalous dimension $\gamma_i$ and of the general form
\begin{equation}
\Sigma_i = a_{i1} [J_0J_0]_s +a_{i2} [J_0J_2]_{s-2}+\cdots,
\end{equation}
with $\langle \Sigma_i \Sigma_j \rangle =\delta_{ij}$. The coefficients $a_{ij}$ are the elements of an orthogonal matrix $O$, with $O O^T=1$, and they generically start at order one in the large $N$ limit. 

Next, we consider four-point functions of the form $\langle J_{s_1} J_{s_2} J_{s_3}J_{s_4}\rangle$, at order $1/N$ (which is the leading connected contribution). By looking at the OPE channel where $J_{s_1}$ approaches $J_{s_2}$, and looking at the logarithmic term in the contribution of operators of twist $2$ and spin $s$, we can read off the following combination of anomalous dimensions :
\begin{equation}\label{combanom}
\sum_i \langle J_{s_1} J_{s_2}  \Sigma_i \rangle \langle \Sigma_i J_{s_3}J_{s_4}\rangle \gamma_i.
\end{equation}

In the large $N$ limit the OPE coefficient $\langle J_{s_1} J_{s_2} [J_{s_5}J_{s_6}]_\ell \rangle$ involves only contractions of $J_{s_1}$ with $J_{s_5}$ and $J_{s_2}$ with $J_{s_6}$ (or the other way around). But this is proportional to $\delta_{s_1,s_5} \delta_{s_2, s_6}$ (since the derivative $\partial^\ell$ consists in having a descendant of these currents). So the coefficients of $\gamma_i$ in \eqref{combanom} are products of the corresponding elements of $a_{ij}$. 

Now we can use some specific results about the correlation functions in the $\phi^6$ theory. At order $1/N$ the four-point function $\langle J_0 J_0 J_0 J_0 \rangle$ is the free theory result plus contributions of order $\lambda_6$ and of order $\lambda_6^2$; more precisely the correction just comes from the $J_0$ exchange diagram, so it is proportional to $\langle J_0 J_0 J_0 \rangle^2 - \langle J_0 J_0 J_0 \rangle_{free}^2 = (a + b \lambda_6)^2 - a^2 = 2 a b \lambda_6 + b^2 \lambda_6^2$ for some constants $a$ and $b$. 
On the other hand, the four-point functions $\langle J_0 J_0 J_{s_1} J_{s_2} \rangle$ (with $s_1,s_2=0,\cdots,\infty$ not both vanishing) are equal at this order to the free theory result plus a contribution of order $\lambda_6$, and all other four-point functions have no corrections. So if we look at the contributions proportional to $\lambda_6^2$ we have $\sum_i a_{i1} a_{i1} \gamma_i$ non-zero, and all the other $\sum_i a_{ik} a_{ij} \gamma_i$ vanish, or in other words $O^T \Gamma O$ has only its $(11)$ element non-vanishing, where the matrix $\Gamma$ is the (diagonal) matrix of anomalous dimensions in the $\Sigma_i$ basis. So this means that in our original basis, at this order, $[J_0 J_0]_s$ is an eigenvector of the anomalous dimension matrix with an eigenvalue proportional to $\lambda_6^2$, and all other eigenvalues vanish at this order in $\lambda_6$.
In particular there is no mixing of this operator with any others at this order, so  in the four-point correlator $\langle J_0 J_0J_0J_0 \rangle$ only a single eigenstate (with twist two and spin $s$) propagates, and its anomalous dimension at order $\lambda_6^2$ is exactly what we read off from this correlation function at order $1/N$.

We can also apply a similar logic to contributions of order $\lambda_6$ (which don't come from $\langle J_0 J_0 J_0 \rangle^2$). However, now the $[J_0 J_{s_1}]_{s-s_1}$ ($s_1 = 0, 2, \cdots, s$) operators can mix with each other and get anomalous dimensions at this order (though other operators still cannot mix with them); such a mixing is exhibited in the explicit results for the 4-point functions $\langle J_0 J_0 J_0 J_s \rangle$ in \cite{Sleight:2018epi}. So there is generally still some non-zero mixing that we cannot resolve without additional information. Thus, we can obtain in this theory exact results about the anomalous dimension matrix at order $\lambda_6^2$, but to go to order $\lambda_6$ requires additional correlation functions.

\section{Double discontinuity sums}
\label{appsums}

In the body of the paper we are led to consider the sum \eqref{gtwosum} of 3d conformal blocks 
\begin{equation}
S(u,v)= \sum_{s=0}^{\infty} a_{0,s}^{(0)} (\gamma_{0,s}^{(1)})^2 f_{\tau=2,s}(u,v).
\end{equation}
This is a generic problem that arises when computing the double-discontinuity due to a tower of operators in the dual channel. In any number of dimensions, and $d=3$ in particular, conformal blocks admit a decomposition in terms of $SL(2,R)$ conformal blocks,
\begin{equation}
f_{\tau,s}(z,\bar z) =\sum_{n=0}^{\infty}  z^{\tau/2+n} \sum_{m=-n}^n c^{(n)}_{m} k_{\tau/2+m+s}(\bar z) ,
\end{equation}
where in the usual normalisations $c^{(0)}_{0}=1$ and the rest of the coefficients can be found by solving the Casimir equation order by order in $z$. At each order in $z$ the generic sums we are interested in take the form
\begin{equation}
S(\bar z) = \sum_{s=0,2,\cdots}  R(s) r_{\tau/2+s} r_{\tau/2} k_{\tau/2+n+s}(\bar z)
\end{equation}
for some integer $n$, where $R(s)$ is some rational function of $s$. Similar sums were considered in \cite{Simmons-Duffin:2016wlq}, but the sums we are interested in do not fall into that category. For the present paper, we are interested in particular in $\tau=2$, but the technique we will apply will be useful in more generality. We will rely on the fact that $SL(2,R)$ conformal blocks are eigenfunctions of the following Casimir operator:
\begin{equation}
\bar D k_{\beta}(\bar z) = \beta(\beta-1) k_{\beta}(\bar z),~~~~~~\bar D =\bar z^2 \partial_{\bar z}(1-\bar z) \partial_{\bar z}.
\end{equation}

Let us assume that we know how to compute the sum
\begin{equation}
h(\bar z) \equiv \sum_{s=0,2,\cdots}  R(s) (4 J^2+1-4k^2) r_{\tau/2+s} r_{\tau/2} k_{\tau/2+n+s}(\bar z),
\end{equation}
where $J^2=(\tau/2+n+s)(\tau/2+n+s-1)$ and $k$ is some constant. We then obtain the following differential equation for $S(\bar z)$,
\begin{equation}\label{eqfors}
(4 \bar D+1-4k^2) S(\bar z) = h(\bar z).
\end{equation}
This equation can be solved by standard Green's function techniques. The homogeneous equation admits two linearly independent solutions
\begin{equation}
\psi_0(\bar z) \equiv \bar z^{\frac{1}{2} (2 k+1)} \, _2F_1\left(k+\frac{1}{2},k+\frac{1}{2};2 k+1;\bar z\right),~~~\psi_1(\bar z) \equiv \bar z^{\frac{1}{2}-k} \, _2F_1\left(\frac{1}{2}-k,\frac{1}{2}-k;1;1-\bar z\right).
\end{equation}
The solution to \eqref{eqfors} is then given by
\begin{equation}\label{eqtwopsis}
S(\bar z) = \frac{\sqrt{\pi } 4^{-k-1} \Gamma \left(k+\frac{1}{2}\right)}{\Gamma (k+1)}\int_{0}^{\bar z} \frac{d\zeta}{\zeta^2}\left( \psi_0(\bar z) \psi_1(\zeta)  - \psi_0(\zeta) \psi_1(\bar z) \right) h(\zeta),
\end{equation}
where the constant of integration has been chosen to get the correct behaviour around $\bar z=0$, and for the applications of this paper $h(\zeta)$ is such that the above integral is convergent. In general this is just a formal solution, as sometimes the resulting integrals are very hard to perform. 

Nevertheless, let us apply this idea to the specific sum \eqref{gtwosum} we need to compute, to leading order in $z$, which takes the form
\begin{equation}
\hat S(\bar z) = \sum_{s=0,2,\cdots}  \frac{r_{1+s} r_{1}}{2s+1} k_{s+1}(\bar z) .
\end{equation}
Instead, we consider the alternative sum
\begin{equation}
\sum_{s=0,2,\cdots} r_{1+s} r_{1}(2s+1) k_{s+1}(\bar z) = \frac{\bar z(2-\bar z)}{4(1-\bar z)},
\end{equation}
as can be shown by using the integral representation for the hypergeometric function, summing over $s$, and then integrating. Our discussion above (for $k=0$) implies that we can write down the result for the sum $\hat S(\bar z)$ as follows:
\begin{equation}\label{repforshat}
\hat S(\bar z) = \frac{\sqrt{\bar z}}{\pi} \int_0^{\bar z} \frac{d\zeta}{\zeta^{3/2}} \left(K(\bar z) K(1-\zeta)-K(\zeta)K(1-\bar z) \right)  \frac{\zeta(2-\zeta)}{2(1-\zeta)},
\end{equation}
where $K(\bar z)$ is the complete elliptic integral of the first kind. It seems that we have not gained much: we could have used a standard integral representation for the hypergeometric function in \eqref{eqtwopsis}, and, after summing over the spin, we would have obtained another integral representation for $\hat S(\bar z)$. The point is that the current representation \eqref{repforshat} makes it much easier to study $\hat S(\bar z)$ around $\bar z=1$. In order to do so, we start with the integrals
\begin{eqnarray}
\int_0^{\bar z} \frac{d\zeta}{\zeta^{1/2}} K(1-\zeta)\frac{(2-\zeta)}{2(1-\zeta)} &=&-\frac{1}{4} \pi \log(1-\bar z) +\frac{1}{4} \pi  (\pi +4\log 2 )-\frac{7\pi}{16}(1-\bar z) + \cdots \\
\int_0^{\bar z} \frac{d\zeta}{\zeta^{1/2}} K(\zeta)\frac{(2-\zeta)}{2(1-\zeta)} &=&\frac{1}{8}\log(1-\bar z)( \log(1-\bar z)- 8 \log 2)+2 C+\frac{\pi ^2}{24}+2 \log^2 2 +\cdots \nonumber
\end{eqnarray}
where $C$ is the Catalan constant. This result can be obtained by studying the integrand around $s=1$, up to a constant term that can be fixed by evaluating the integrals at $\bar z=1$, after subtracting the divergent contribution. Having these results, we can now write down 
\begin{eqnarray}\label{finalshat}
\hat S(\bar z) = \frac{1}{16} \log^2(1-\bar z) -\frac{1}{8} (\pi +4 \log 2) \log(1-\bar z)-C-\frac{\pi ^2}{48}+\log ^2 2+\frac{1}{2} \pi  \log 2+ \cdots.
\end{eqnarray}

Now we would like to discuss a simpler, although less rigorous, way to obtain these results. Let us consider $\hat S(\bar z)$ again, and assume we can expand each conformal block around $\bar z=1$, before performing the sum. Keeping only the first order we obtain
\begin{eqnarray}
\hat S(\bar z) = \sum_{s=0,2,\cdots} \left(- \frac{2 H(s)}{2s+1} - \frac{\log (1-\bar z)}{2s+1} \right),
\end{eqnarray}
where $H(s)$ is the harmonic number. The sums over $s$ are now divergent. This is a manifestation of the fact that $\hat S(\bar z)$ should have an enhanced singularity, and it is not correct to expand around $\bar z=1$ before doing the sum. Nevertheless, let us regularise the above sums by adding a factor $e^{-s \epsilon}$. Performing the sums and expanding around $\epsilon=0$ we obtain
\begin{eqnarray}
\sum_{s=0,2,\cdots} \frac{e^{-s \epsilon}}{2s+1} &=& \frac{1}{8} (-2 \log \epsilon+\pi +4 \log2)+ \cdots,\\
\sum_{s=0,2,\cdots}  \frac{H(s)}{2s+1}e^{-s \epsilon} &=&\frac{\log^2 \epsilon}{8}+\frac{C}{2}+\frac{\pi ^2}{24}-\frac{1}{2} \log ^2 2-\frac{1}{4} \pi  \log 2+ \cdots.
\end{eqnarray}
We see that the enhanced singularity as $v \to 0$ arises as a divergence as $\epsilon \to 0$. Furthermore, we expect $\epsilon$ and $v$ to be related as follows. The factor $e^{-s \epsilon}$ damps the sum for spins of order $s \epsilon \sim 1$. From the analysis of \cite{Alday:2007mf}, it follows that the spin should be related to the cross ratio as $ s \sim 1/\sqrt{v}$. This implies $\epsilon \sim \sqrt{1-\bar z}$. The overall coefficient, as well as higher order terms, do not follow easily from the analysis of \cite{Alday:2007mf}. The precise relation appears to be
\begin{eqnarray}
\log \epsilon = \frac{1}{2} \log (1-\bar z) + i \frac{\pi}{2} + \cdots.
\end{eqnarray}
With this, the regularised sum exactly agrees with the correct result! This method allows the corresponding sums to be evaluated to any desired order in $z$. It can be checked numerically, with a very high precision, that the correct result \eqref{finalshat} is reproduced. 

\bibliographystyle{utphys} 
\bibliography{CSbootstrap}

\end{document}